\documentclass{sig-alternate}
\usepackage[ruled,linesnumbered,noline,noend]{algorithm2e}
\usepackage{enumitem}
\usepackage{balance}

\SetVlineSkip{0pt}   
\begin{document}

%Alinhamento do texto... Site: http://sumanta679.wordpress.com/2009/05/20/latex-justify-without-hyphenation/
\tolerance=1000
 \emergencystretch=\maxdimen
 \hyphenpenalty=1
 \hbadness=1
%%%%%%%%%%%%%%%%%%%%%%%%%%%%%%%%%%%%%%	

%
% --- Author Metadata here ---
\conferenceinfo{SAC'15}{April 13-17, 2015, Salamanca, Spain}
\CopyrightYear{2015}
\copyrightetc{Copyright is held by the owner/author(s). Publication rights licensed to ACM. \\ ACM \the\acmcopyr}
\crdata{978-1-4503-3196-8/15/04...\$15.00.\\
http://dx.doi.org/10.1145/2695664.2695850}
% --- End of Author Metadata ---

%\title{Using GPU to Parallelize Recommender Systems}
\title{Accelerating Recommender Systems using GPUs}
%\subtitle{[Extended Abstract]
%\titlenote{A full version of this paper is available as
%\textit{Author's Guide to Preparing ACM SIG Proceedings Using
%\LaTeX$2_\epsilon$\ and BibTeX} at

\numberofauthors{3} %  in this sample file, there are a *total*
\author{
% 1st. author
 \alignauthor
 Andr\'e Valente Rodrigues\\%\titlenote{Researcher.}\\
 			 \affaddr{LIAAD - INESC TEC}\\
        \affaddr{DCC - University of Porto}
 % 2nd. author
 \alignauthor Al\'ipio Jorge\\%\titlenote{Supervisor on recommender systems}\\
 			 \affaddr{LIAAD - INESC TEC}\\
        \affaddr{DCC - University of Porto}
 % 3rd. author
 \alignauthor In\^es Dutra\\%\titlenote{Supervisor on parallel systems}\\
 			 \affaddr{CRACS - INESC TEC}\\
        \affaddr{DCC - University of Porto}
 %\and  % use '\and' if you need 'another row' of author names
 }
% There's nothing stopping you putting the seventh, eighth, etc.
% author on the opening page (as the 'third row') but we ask,
% for aesthetic reasons that you place these 'additional authors'
% in the \additional authors block, viz.
%\additionalauthors{Additional authors: John Smith (The
%Th{\o}rv{\"a}ld Group, email: {\texttt{jsmith@affiliation.org}})
%and Julius P.~Kumquat (The Kumquat Consortium, email:
%{\texttt{jpkumquat@consortium.net}}).}
% Just remember to make sure that the TOTAL number of authors
% is the number that will appear on the first page PLUS the
% number that will appear in the \additionalauthors section.

\maketitle
\begin{abstract}
We describe GPU implementations of the matrix recommender algorithms
CCD++ and ALS. We compare the processing time and predictive ability
of the GPU implementations with existing multi-core versions of the
same algorithms. Results on the GPU are better than the results
of the multi-core versions (maximum speedup of 14.8). 
%GPU-CCD++ is also faster than GPU-ALS.
%% We parallelize the CCD++ algorithm for GPUs and compare our results
%% with the ones obtained with a multicore implementation and with
%% ALS. Results show that we can obtain very good speedups (14.8 maximum)
%% on the GPUs, better than the results obtained with the multicore
%% version and better than ALS.
 %\textbf{\underline{(Acrescentar que o ALS foi
            %implementado)}}  
\end{abstract}


\section{Introduction}
Recommendation (or Recommender) systems are capable of predicting
user responses to a large set of
options~\cite{Burke:2007:HWR:1768197.1768211,Mahmood:2009:IRS:1557914.1557930,Resnick:1997:RS:245108.245121}.
They are generally implemented in web site applications related with
music, video, shops, among others, and collect information about
preferences of different users in order to predict the
next preferences. More recently, social network sites such as
Facebook, also started to use recommender algorithms~\cite{Baatarjav:2008:GRS:1484422.1484475-reduced,He:2010:SNR:2049515}.

Many recommendation systems are implemented using matrix factorization
algorithms~\cite{Koren_Handbook_p145,Sarwar00applicationof}. Given a
user-item interaction matrix $A$, the objective of these algorithms is
to find two matrices $W$ and $H$ such that $W \times H^T$ approximate
$A$. Matrix $W$ represents user profiles and $H$ represents
items. With $W$ and $H$ we can easily predict the preference of user
$i$ regarding an item $j$. The fact that these
recommendation systems are based on matrices operations make them
suitable to parallelization. In fact, due to the huge sizes of $W$ and
$H$ and the nature of these algorithms, many authors have pursued the
parallelization path. For example, popular algorithms like the
Alternating Least Squares (ALS), or the Stochastic Gradient Descent
(SGD) have parallel versions for either shared-memory or distributed
memory
architectures~\cite{Takacs:2012:ALS:2365952.2365972,kais-pmf,6153051,Zinkevich10parallelizedstochastic}.
Recently, Yu {\it et al.}~\cite{kais-pmf} demonstrated that coordinate
descent based methods (CCD) have a more efficient update rule compared
to ALS. They also show more stable convergence than SGD. They
implemented a new recommendation algorithm using CCD as the basic
factorization method and showed that CCD++ is faster than both
SGD and ALS.

%factorization method and showed that CCD++ is 4 times faster than both
%SGD and ALS using a distributed system with 20 nodes.
%factorization method and showed that CCD++ is 40 times faster than SGD and 20 times faster than ALS using a distributed system with 256 nodes.
With the increasing popularity of general purpose graphics
processing units (GPGPU), and their suitability to data parallel
programming, algorithms that are based on data matrices operations
have been successfully deployed to these platforms, taking advantage
of their hundreds of cores. Regarding recommendation systems, we are
only aware of the work of Zhanchun {\it et al.}~\cite{gao:improving} that
implemented a neighborhood-based algorithm for GPUs.
% é preciso explicar o que faz o gao e porque o nosso ainda assim inova.
In this paper, we describe GPU implementations of two recommendation algorithms based on matrix factorization. This is, to our knowledge, the first proposal of this kind in the field.

We implement CCD++ and ALS GPU versions using the CUDA programming
model in Windows and Linux.  We tested our versions on typical
benchmarks found in the literature.  We compare our results with an
existing multi-core version of CCD++ and our own multi-core
implementation of ALS. Our best results with GPU-CCD++ and GPU-ALS
versions show speedups of 14.8 and 6.2, respectively over their
sequential versions (single core).  The fastest CUDA version (CCD++ on
windows) is faster than the fastest 32-core version.  All results
on the GPU and multi-core have the same recommendation quality (same root mean
squared error) as the sequential implementations.

%We also show that
GPU-CCD++ can be a better parallelization choice over a multi-core
implementation, given that it is much cheaper to buy a machine with a
GPGPU with hundreds of cores than to buy a multi-core machine with a
few cores.  

Next, we present basic concepts about GPU programming and
architecture, explain the basics of factorization algorithms and their
potential to parallelization, describe our own parallel implementation
of ALS and CCD++, show results of experiments performed with
typical benchmark data and, finally, we draw some
conclusions and perspectives of future work.

\section{The CUDA Programming Model}

The CUDA programming model~\cite{hochberg2012matrix,Sanders:2010:CEI:1891996,wilt2013cuda}, developed by NVIDIA,
is a platform for parallel computing on Graphics Processing Units
(GPU). One single host machine can have one or multiple GPUs, having a
very high potential for parallel processing. GPUs were mainly designed
and used for graphics processing tasks, but currently, with tools
like CUDA or OpenCL~\cite{Fang:2011:CPC:2066302.2066955}, other kinds of applications can take
advantage of the many cores that a GPU can provide.  This motivated
the design of what today is called a GPGPU (General Purpose
  Graphics Processing Unit), a GPU for multiple
purposes~\cite{hochberg2012matrix,Sanders:2010:CEI:1891996,wilt2013cuda}. 

GPUs fall in to the single-instruction-multiple-data (SIMD)
architecture category, where many processing elements simultaneously
run the {\em same program} but on distinct data items.  This program,
referred to as the {\em kernel}, can be quite complex including
control statements such as {\em if} and {\em while}.

Scheduling work for the GPU is as follows. A thread in the host platform
(e.g., a multi-core) first copies the data to be processed from host
memory to GPU memory, and then invokes GPU threads to run the {\em
  kernel} to process the data. Each GPU thread has a unique id which
is used by each thread to identify what part of the data set it will
process. When all GPU threads finish their work, the GPU signals the
host thread which will copy the results back from GPU memory to host
memory and schedule new work~\cite{wilt2013cuda}.%~\cite{hlpp2014}.

GPU memory is organized hierarchically and each (GPU) thread has its own {\em
  per-thread local} memory. Threads are grouped into {\em blocks},
each {\em block} having a memory {\em shared} by all threads in the {\em block}.
Finally, thread {\em blocks} are grouped into a single {\em grid} to execute
a {\em kernel} --- different {\em grids} can be used to run different {\em kernels}.
All {\em grids} share the {\em global memory}.

All data transfers between the host (CPU) and the GPU are made through
reading and writing global memory, which is the slowest. A
common technique to reduce the number of reads from global memory is
{\em coalesced memory access}, which takes place when consecutive
threads read consecutive memory locations allowing the hardware to
coalesce the reads into a single one.

Programming a GPGPU is not a trivial task when algorithms do not
present regular computational patterns when accessing data. But a GPU
brings a great advantage over multiprocessors, since it has hundreds
of processing units that can perform data parallelism, present in many
applications, specially the ones that are based on recommendation
algorithms, and because it is much cheaper than a (CPU) with
a few cores. 

\section{Matrix Factorization}

%ID: Andr�, qual � a diferen�a entre collaborative filtering e matrix
%factorization? A ref que coloquei acima implementa collaborative
%filtering em GPUs. Neste caso, qual � a diferen�a em rela��o ao CCD++?
Collaborative filtering Recommendation algorithms can be implemented
using different techniques, such as neighborhood-based and association
rules. One powerful family of
collaborative filtering algorithms use another technique known as matrix
factorization ~\cite{Koren_Handbook_p145,Sarwar00applicationof}
resourcing to the UV-Decomposition or SVD (Singular Value
  Decomposition) matrix factorization methods.  SVD is also commonly
used for image and video
compression~\cite{1162766,isbnplus9780321796974,Koren_Handbook_p145,Sarwar00applicationof}.

\begin{figure}[htb]
	\centering
	\includegraphics[width=0.3\textwidth]{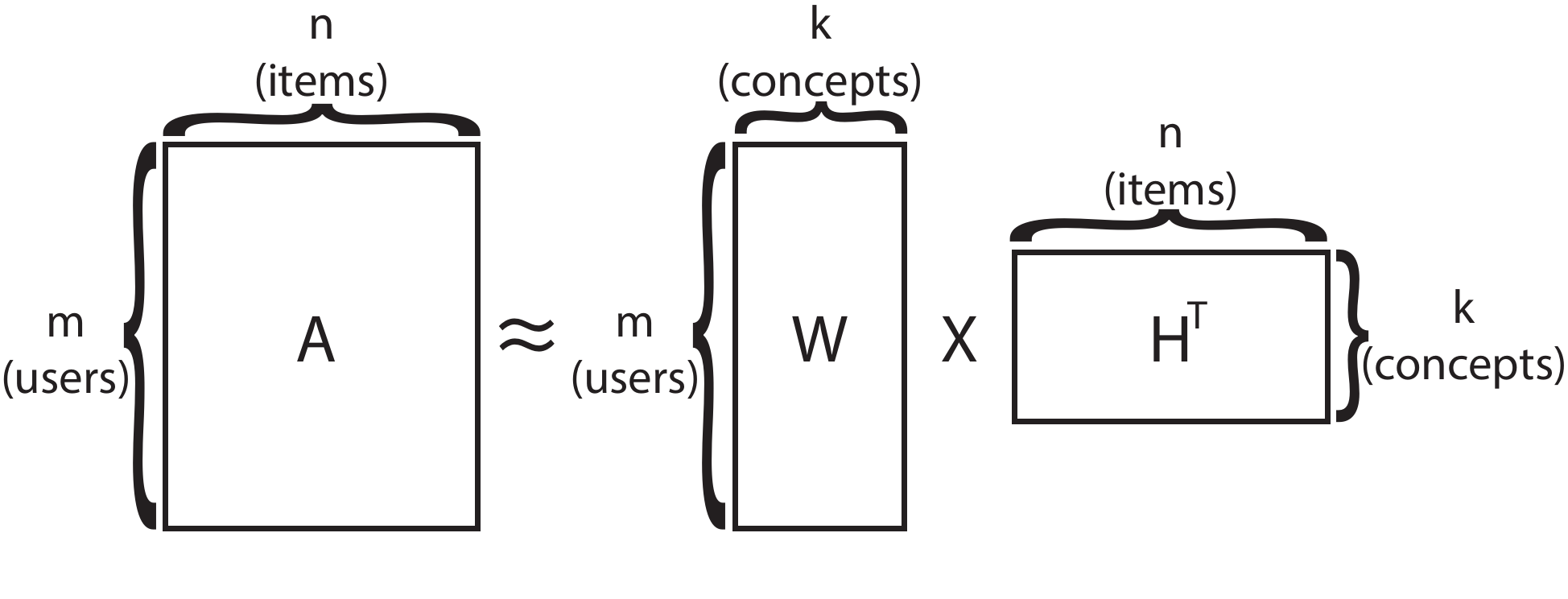}
	\caption{Obtaining $A$.}
	\label{fig:uvDecomp}
	\end{figure}

The UV-Decomposition approach is applied to learning a
recommendation model as follows. Matrix $A$ is the $m \times n$
ratings matrix and contains a non-zero $A_{i,j}$ value for each (user
$i$)--(item $j$) interaction. Using a matrix factorization (MF) algorithm,
we obtain matrices $W\in\mathbb{R}^{m\times k}$ and
$H\in\mathbb{R}^{n\times k}$ whose product approximates $A$
(Figure~\ref{fig:uvDecomp}). The matrix $W$ profiles the users using
$k$ latent features, known as factors. The matrix $H$ profiles the
items using the same features. By the nature of the recommendation
problem, $A$ is a sparse matrix that contains mostly zeros (user-item
pairs without any interaction). In fact, this matrix is never
explicitly represented, but we can estimate any of its unknown values
$A_{i,j}$ by computing the dot product of row $i$ of $W$ and row $j$
of $H$.  With these estimated values we can produce recommendations.
	
The matrices $W$ and $H$ are obtained by minimizing the objective
function in eq.~(\ref{eq:MainObjectiveFunction}). In this function, $A\in
\mathbb{R}^{m\times n}$ is the classification matrix, $m$ is the
number of users and $n$ is the number of items.
	
	\begin{equation} \label{eq:MainObjectiveFunction}
		\underset{H \in \mathbb{R}^{n\times k}}{\underset{W \in \mathbb{R}^{m\times k}}{\text{min}}} \sum\limits_{(i,j) \in \Omega} (A_{ij}-\omega^T_i h_j)^2 + \lambda (\lvert \lvert W \rvert \rvert ^2_F + \lvert \lvert H \rvert \rvert ^2_F),
	\end{equation}	

\noindent	
Assuming that the classification matrix is sparse (i.e., a minority of ratings is known),  $\Omega$ is the set of indexes related to the
observed classifications (ratings), $i$ is the user counter and $j$ is
the item counter. The sparse data is represented by the triplet $i, j,
classification$. The
$\lambda$ parameter is a regularization factor, which determines how
precise will be the factorization given by the objective function. 
In other words, it allows to control the error level and 
overfitting. The Frobenius norm indicated by $\lvert
\lvert \star \rvert \rvert _F$, is used to calculate the distance
between the matrix $A$ and its approximate matrix $rank-k = A_k$. In
this context, $E=A-A_k$, is the Frobenius norm, which consists
of calculating
$\lvert \lvert E \rvert \rvert _F^2 = \sum_{i,j}\lvert E_{i,j} \rvert
^2$. The lower the integer produced by the summation, the nearer is
$A$ to $A_k$ ~\cite{Meyer:2000:MAA:343374}. The 
$\omega^T_i$ vector corresponds to line $i$ of matrix $W$ and the
$h_j$ vector corresponds to the line $j$ of matrix $H$. Summarizing,
the objective function is used to obtain an approximation of the
incomplete matrix $A$, where $W$ and $H$ are matrices $rank-k$.

It is not trivial to directly calculate the minimum of the objective
function in eq.~(\ref{eq:MainObjectiveFunction}). Therefore, to solve the
problem, several methods are used. Next, we explain some of them, most
relevant to this work.
	
\subsection{Alternating Least Squares (ALS)}
This method divides the minimization function in two quadratic
functions. That way, it minimizes $W$ keeping $H$ constant and it
minimizes $H$ keeping $W$ constant. When $H$ is constant to minimize
$W$, in order to obtain an optimal value to $\omega _i^*$, the
function in eq.~(\ref{eq:ALS_ObjectiveFunction}) is derived.
	
	\begin{equation} \label{eq:ALS_ObjectiveFunction}
		min_{\omega_i} \sum\limits_{j \in \Omega _i} (A_{ij} - w^T_i h_j)^2 + \lambda \lvert \lvert w_i \rvert \rvert ^2
	\end{equation}
\noindent

Next, it is necessary to minimize
function in eq.~(\ref{eq:ALS_ObjectiveFunction}). The expression:

	\[\omega_i^*= (H_{\Omega _i}^T H_{\Omega _i} + \lambda I )^{-1} H^T a_i\]

\noindent
gives a minimal value for $\omega_i^*$, given that $\lambda$ is always
positive. 

The algorithm alternates between the minimizations of $W$ and $H$
until its convergence, or until it reaches a determined number $T$ of
iterations, given by the
user~\cite{Takacs:2012:ALS:2365952.2365972,kais-pmf,6153051}.

In our implementation, the inverse matrix is obtained using the
Cholesky decomposition, since it is one of the most efficient methods
for matrix inversion~\cite{CholeskyForInv}. 
%A linguagem de implementa��o � \emph{C++}.

The complete sequential ALS is shown in Algorithm~\ref{alg:ALS_ALG}.

{\footnotesize
\SetAlFnt{\small}
%\IncMargin{1em}
\begin{algorithm}[htb]
\SetKwData{Left}{left}\SetKwData{This}{this}\SetKwData{Up}{up}
\SetKwFunction{Union}{Union}\SetKwFunction{FindCompress}{FindCompress}
\SetKwInOut{Input}{input}\SetKwInOut{Output}{output}
\SetKwFunction{Initialize}{Initialize}
\Input{$A, W, H, \lambda, T$}
	\Initialize{$H \leftarrow \text{(small random numbers)}$}\;
	\For{$iter\leftarrow 1$ \KwTo $T$ $Step=1$}{
		Compute the $W$ using $\omega_i^*= (H_{\Omega _i}^T H_{\Omega _i} + \lambda I )^{-1} H^T a_i$\;
		Compute the $H$ using $h_j^*= (W_{\Omega _j}^T W_{\Omega _j} + \lambda I )^{-1} W^T a_j$\;
	}
\caption{ALS~\cite{Takacs:2012:ALS:2365952.2365972}}\label{alg:ALS_ALG}
\end{algorithm}%\DecMargin{1em}
} % end small

\subsection{Cyclic Coordinate Descent (CCD)}
The algorithm is very similar to ALS, but instead of minimizing
function in eq.~(\ref{eq:MainObjectiveFunction}) for all elements of $H$ or
$W$, it minimizes the function for each element of $H$ or $W$ at each
iteration step~\cite{Hsieh:2011:FCD:2020408.2020577-reduced,kais-pmf}.
Assuming $\omega_{i}$ represents the line $i$ of $W$, then
$\omega_{it}$ represents the element of line $i$ and column $t$. In
order to operate element by element, the objective
function in eq.~(\ref{eq:MainObjectiveFunction}) needs to be modified such that
only $\omega_{it}$ can be assigned a $z$ value. This reduces the
problem to a single variable problem, as shown in
function in eq.~(\ref{eq:mainCCDFunc}).
	\begin{equation} \label{eq:mainCCDFunc}
		\underset{z}{\text{min}} f(z) = \sum\limits_{j \in \Omega_i} (A_{ij}-(\omega^T_i h_j-\omega _{it} h_{jt})-zh_{jt})^2 + \lambda z^2,
	\end{equation}
\noindent
Given that this algorithm performs a non-negative matrix factorization
and function in eq.~(\ref{eq:mainCCDFunc}) is invariably quadratic, it has one
single minimum. Therefore, it is sufficient to
minimize function in eq.~(\ref{eq:mainCCDFunc}) in relation to $z$,
obtaining eq.~(\ref{eq:minTo_z_CCD}).
	
	\begin{equation} \label{eq:minTo_z_CCD}
		z^* = \frac{\sum\limits_{j \in \Omega_i} (A_{ij}-\omega^T_i h_j+\omega _{it} h_{jt})h_{jt}}{\lambda + \sum\limits_{j \in \Omega_i} h^2_{jt}}, 
	\end{equation}	
\noindent
Finding $z^*$ requires
$O(\lvert \Omega_i \rvert k)$ iterations. If $k$ is large, this
step can be optimized after the first iteration, thus requiring only
$O(\lvert \Omega_i \rvert)$ iterations. In order to do that, it
suffices to keep a residual matrix $R$
such that $R_{ij} \equiv A_{ij}-\omega^T_i h_j, \forall (i,j) \in
\Omega$. Therefore, after the first iteration, and after obtaining
$R_{ij}$, the minimization of $z*$ becomes:
{\small
	\begin{equation} \label{eq:minTo_z_ResidualR_CCD}
		z^* = \frac{\sum\limits_{j \in \Omega_i} (R_{ij}+\omega _{it} h_{jt})h_{jt}}{\lambda + \sum\limits_{j \in \Omega_i} h^2_{jt}}, 
	\end{equation}	
}
\noindent	
Having calculated $z^*$, the update of $\omega_{it}$ and $R_{ij}$
proceeds as follows:
{\small
	\begin{equation} \label{eq:updateRbyZ}
		R_{ij} \leftarrow R_{ij}-(z^* -\omega_{it})h_{jt}, \forall j \in \Omega_i, 
	\end{equation}	
		\begin{equation} \label{eq:updateOmega}
		\omega _{it} \leftarrow z^*.
	\end{equation}	
}
After updating each variable $\omega_{it} \in W$
using~(\ref{eq:updateOmega}), we need to update the variables $h_{jt}
\in H$ in a similar manner, obtaining:
%where in the first iteration $s^*$ is given by:

%% 	\begin{equation} \label{eq:minTo_s_CCD}
%% 		s^* = \frac{\sum\limits_{i \in \bar{\Omega}_j} (A_{ij}-\omega^T_i h_j+\omega _{it} h_{jt})\omega _{it}}{\lambda + \sum\limits_{i \in \bar{\Omega}_j} \omega ^2_{it}}, 
%% 	\end{equation}		
%% \noindent
%% The residual matrix $R$ is again calculated with $R_{ij} \equiv
%% A_{ij}-\omega^T_i h_j, \forall (i,j) \in \Omega$. Again, after
%% obtaining $R_{ij}$ in the first iteration, the minimization $s*$
%% becomes:
{\small
 	\begin{equation} \label{eq:minTo_s_ResidualR_CCD}
 		s^* = \frac{\sum\limits_{i \in \bar{\Omega}_j} (R_{ij}+\omega _{it} h_{jt})\omega _{it}}{\lambda + \sum\limits_{i \in \bar{\Omega}_j} \omega ^2_{it}}, 
 	\end{equation}
}
%% After $s^*$ is calculated, the update of $h_{jt}$ and $R_{ij}$ is
%% perfomed as follows:
{\small
	\begin{equation} \label{eq:updateRbyS}
		R_{ij} \leftarrow R_{ij}-(s^* -h_{jt})\omega_{it}, \forall i \in \bar{\Omega}_j, 
	\end{equation}	
		\begin{equation} \label{eq:updateH}
		h_{jt} \leftarrow s^*.
	\end{equation}	
}
Having obtained the updating rules shown
in eqs.~(\ref{eq:updateRbyZ}), ~(\ref{eq:updateOmega}), ~(\ref{eq:updateRbyS})
and ~(\ref{eq:updateH}), we can now apply any sequence of updates to $W$
and $H$.  Next, we describe two ways of performing the updates: 
item/user-wise and feature-wise.
	
\subsubsection{Update item/user-wise CCD}	
In this type of updating, $W$ and $H$ are updated as % follows:
in Algorithm~\ref{alg:ccdALG}.

%% \[
%% \overbrace{\underbrace{\omega _{11},\ldots,\omega _{1k}}_{\omega _1},\ldots,
%% 						\underbrace{\omega _{m1},\ldots,\omega _{mk}}_{\omega _m}}
%% ^W
%% ,
%% \overbrace{\underbrace{h _{11},\ldots,h _{1k}}_{h _1},\ldots,
%% 						\underbrace{h _{n1},\ldots,h _{nk}}_{h _n}}
%% ^H
%% .
%%  \]

In the first iteration $W$ is initialized with zeros, therefore the
residual matrix $R$ is exactly equal to
$A$. 
%Algorithm~\ref{alg:ccdALG} shows this procedure, with $T$
%corresponding to the number of iterations.

\SetAlFnt{\small}
{\small
%\IncMargin{1em}
\begin{algorithm}[htb]
\SetKwData{Left}{left}\SetKwData{This}{this}\SetKwData{Up}{up}
\SetKwFunction{Union}{Union}\SetKwFunction{FindCompress}{FindCompress}
\SetKwInOut{Input}{input}\SetKwInOut{Output}{output}
\SetKwFunction{Initialize}{initialize}
\Input{$A, W, H, \lambda, k, T$}
	\Initialize{$W\leftarrow 0, R\leftarrow A$}\;	
	\For{$iter\leftarrow 1$ \KwTo $T$ $Step=1$}{		
		\For(\tcp*[h]{$\rhd$ Update $W$.}){$i\leftarrow 1$ \KwTo $m$ $Step=1$}{%\tcp*[f] right justified...
			\For{$t\leftarrow 1$ \KwTo $k$ $Step=1$}{
				obtain $z^*$ using (\ref{eq:minTo_z_ResidualR_CCD})\;
				update $R$ and $\omega_{it}$ using (\ref{eq:updateRbyZ}) and (\ref{eq:updateOmega})\;
			}		
		}
		\For(\tcp*[h]{$\rhd$ Update $H$.}){$j\leftarrow 1$ \KwTo $n$ $Step=1$}{
			\For{$t\leftarrow 1$ \KwTo $k$ $Step=1$}{
				obtain $s^*$ using (\ref{eq:minTo_s_ResidualR_CCD})\;
				update $R$ and $h_{jt}$ using (\ref{eq:updateRbyS}) and (\ref{eq:updateH})\;
			}		
		}		

	}
\caption{CCD~\cite{kais-pmf}}\label{alg:ccdALG}
\end{algorithm}%\DecMargin{1em}
} % end small

\subsubsection{Update feature-wise CCD++}	

Assuming that $\bar{\omega}_t$ corresponds to the columns of $W$ and
$\bar{h}_t$, the columns of $H$, the factorization $WH^T$ can be represented as a summation of $k$ outer products.
{\small
	\begin{equation} \label{eq:CCDplusplus_newRep}
		A \approx WH^T = \sum\limits_{t=1}^k \bar{\omega}_t \bar{h}_t^T, 
	\end{equation}	
}	
\noindent
%% com esta conclus�o, em vez da atualiza��o ser de item e item $m$ ou
%% utilizado em utilizador $n$ alternando entre $H$ e $W$, passa a ser de
%% caracter�stica latente em caracter�stica latente $k$ entre as $H$ e
%% $W$ simultaneamente, da� $\bar{\omega}_t \in \mathbb{N}^m$, $\bar{h}_t
%% \in \mathbb{N}^n$ e $t\in \mathbb{N}^k$.

Some modifications need to be made to the original CCD
functions. Assuming that $u^*$ and
$v^*$ are the vectors to be injected over
$\bar{\omega}_t$ and $\bar{h}_t$, then $u^*$ and $v^*$ can be
calculated using the following minimization:
{\small
	\begin{equation} \label{eq:mainCCDFunc_feauture}
		\underset{u\in \mathbb{R}^m,v\in \mathbb{R}^n}{\text{min}} \sum\limits_{(i,j) \in \Omega} (R_{ij} + \bar{\omega}_{ti} \bar{h}_{tj} - u_i v_j)^2 + \lambda (\lvert \lvert u \rvert \rvert ^2 + \lvert \lvert v \rvert \rvert ^2),
	\end{equation}	
}
\noindent

$R_{ij} \equiv A_{ij}-\omega^T_i h_j, \forall (i,j) \in \Omega$ is the
residual entry of $(i,j)$. But using this type of update, there is one
more possibility which is to have pre-calculated values using a second
residual matrix $\hat{R}_{ij}$:
{\small	
	\begin{equation} \label{eq:computeCCD_feautureR}
		\hat{R}_{ij} = R_{ij}+\bar{\omega} _{ti}\bar{h} _{tj}, \forall (i,j) \in \Omega ,
	\end{equation}		
}
\noindent
This way, the objective function equivalent to
(\ref{eq:MainObjectiveFunction}) is rewritten as:
{\small
	\begin{equation} \label{eq:mainCCDFunc_feautureFinal}
		\underset{u\in \mathbb{R}^m,v\in \mathbb{R}^n}{\text{min}} \sum\limits_{(i,j) \in \Omega} (\hat{R}_{ij} - u_i v_j)^2 + \lambda (\lvert \lvert u \rvert \rvert ^2 + \lvert \lvert v \rvert \rvert ^2).		
	\end{equation}		
}
To obtain $u^*$ it suffices to minimize the function (\ref{eq:mainCCDFunc_feautureFinal}) regarding $u_i$:
{\small	
	\begin{equation} \label{eq:minUi_CCDFeauture}
		u_i \leftarrow \frac{\sum\limits_{j \in \Omega_i} \hat{R}_{ij} v_j}{\lambda + \sum\limits_{j \in \Omega_i} v^2_j}, i=1,\ldots ,m,
	\end{equation}
}
\noindent
To obtain $v^*$ it suffices to minimize
(\ref{eq:mainCCDFunc_feautureFinal}) regarding $v_j$:
{\small
	\begin{equation} \label{eq:minVj_CCDFeauture}
		v_j \leftarrow \frac{\sum\limits_{i \in \bar{\Omega}_j} \hat{R}_{ij}u_j}{\lambda + \sum\limits_{i \in \bar{\Omega}_j} u^2_i}, j=1,\ldots ,n.
	\end{equation}		
}	
%% Note-se que devido � independ�ncia dos vetores $u^*$ e $v^*$ em
%% rela��o �s restantes vari�veis, por isso para cada $k$, $u^*$ e $v^*$
%% apenas dependem um do outro para atingirem a m�xima otimiza��o, logo
%% com esta assun��o, ap�s o ambiente estar criado por $\hat{R}_{ij}$,
%% basta apenas aplicar \emph{CCD} para otimizar as posi��es dos vetores
%% $u^*$ e $v^*$ alternando entre um e outro tantas vezes quantas as
%% necess�rias, sendo que cada conjunto $u^*$ e $v^*$ apenas � visitado
%% uma �nica vez pelas itera��es \emph{CCD}, em que segundo
%% ~\cite{kais-pmf} uma boa forma de inicializar $(u^*,v^*)$ � utilizar
%% $(\bar{\omega}_t,\bar{h}_t)$.
	
Finally, after obtaining $u^*$ e $v^*$ we update
$(\bar{\omega}_t,\bar{h}_t)$ and $R_{ij}$:
{\small
	\begin{equation} \label{eq:updateOmegaH_CCDPP}
		(\bar{\omega}_t,\bar{h}_t) \leftarrow (u^*,v^*),
	\end{equation}	

	\begin{equation} \label{eq:updateR_CCDPP}
		R_{ij} \leftarrow \hat{R}_{ij} - u_i^* v_j^*, \forall (i,j) \in \Omega ,
	\end{equation}					
}
Algorithm~\ref{alg:ccdppALG} formalizes the feature-wise update of
CCD, called CCD++.

{\small	
%\IncMargin{1em}
\SetAlFnt{\small}
\begin{algorithm}[htb]
\SetKwData{Left}{left}\SetKwData{This}{this}\SetKwData{Up}{up}
\SetKwFunction{Union}{Union}\SetKwFunction{FindCompress}{FindCompress}
\SetKwInOut{Input}{input}\SetKwInOut{Output}{output}
\SetKwFunction{Initialize}{initialize}
\Input{$A, W, H, \lambda, k, T$}
	\Initialize{$W\leftarrow 0, R\leftarrow A$}\;	
	\For{$iter\leftarrow 1$ $\ldots$ $Step=1$}{		
		\For{$t\leftarrow 1$ \KwTo $k$ $Step=1$}{
			build $\hat{R}$ using (\ref{eq:computeCCD_feautureR})\;
			\For(\tcp*[h]{$\rhd$ $T$ iterations CCD to (\ref{eq:mainCCDFunc_feautureFinal}).}){$inneriter\leftarrow 1$ \KwTo $T$ $Step=1$}{				
				update $u$ using (\ref{eq:minUi_CCDFeauture})\;
				update $v$ using (\ref{eq:minVj_CCDFeauture})\;
			}	
			update $(\bar{\omega}_t,\bar{h}_t)$ and $R$ using (\ref{eq:updateOmegaH_CCDPP}) and (\ref{eq:updateR_CCDPP})\;		
		}	

	}
\caption{CCD++~\cite{kais-pmf}}\label{alg:ccdppALG}
\end{algorithm}%\DecMargin{1em}
} % end small

%\newpage

\section{Parallel ALS}
Parallelizing ALS consists of distributing the matrices $W$ and $H$ among
threads. Synchronization is needed as soon as the matrices are updated
in
parallel~\cite{Zhou08large-scaleparallel}. Algorithm~\ref{alg:ALSP_ALG}
shows the modifications related to the sequential ALS algorithm.
{\small
%\IncMargin{1em}
\SetAlFnt{\small}
\begin{algorithm}[htb]
\SetKwData{Left}{left}\SetKwData{This}{this}\SetKwData{Up}{up}
\SetKwFunction{Union}{Union}\SetKwFunction{FindCompress}{FindCompress}
\SetKwInOut{Input}{input}\SetKwInOut{Output}{output}
\SetKwFunction{Initialize}{Initialize}
\Input{$A, W, H, \lambda, T$}
\Begin{
	\Initialize{$H \leftarrow \text{(small random numbers)}$}\;
	\For{$iter\leftarrow 1$ \KwTo $T$ $Step=1$}{
		Compute in parallel the $W$ using $\omega_i^*= (H_{\Omega _i}^T H_{\Omega _i} + \lambda I )^{-1} H^T a_i$;		
		\tcp*[h]{Sync}\;
		Compute in parallel the $H$ using $h_j^*= (W_{\Omega _j}^T W_{\Omega _j} + \lambda I )^{-1} W^T a_j$;
		\tcp*[h]{Sync}\;
	}
}
\caption{Parallel ALS}\label{alg:ALSP_ALG}
\end{algorithm}%\DecMargin{1em}
} %end small

%% Sabendo que o algoritmo otimiza as matrizes $W$ e $H$ linha a linha, e
%% para otimizar $W$, $H$ � mantida fixa, ent�o as linhas de $W$ podem
%% ser calculadas de forma independente. O mesmo acontece quanto
%% otimiza-se $H$, mantendo $W$ fixa.
	
%In our parallel version we use the \emph{OpenMP}
%library~\cite{chandra2001parallel}.  

\section{GPU-ALS}
We also parallelized ALS for CUDA. Data are copied to the GPU
and the host is responsible for the synchronization. When the
computation finishes in the GPU, $W$ and $H$ are copied from the
device to the host. Algorithm~\ref{alg:ALSP_CUDA_ALG} shows how ALS
was parallelized using CUDA.
{\small
%\IncMargin{1em}	
\SetAlFnt{\small}
\begin{algorithm}[htb]
\SetKwData{Left}{left}\SetKwData{This}{this}\SetKwData{Up}{up}
\SetKwFunction{Union}{Union}\SetKwFunction{FindCompress}{FindCompress}
\SetKwInOut{Input}{input}\SetKwInOut{Output}{output}
\SetKwFunction{Initialize}{Intialize}
\Input{$A, W, H, \lambda, T$}
Allocate GPU memory for matrices $A$, $W$ and $H$\;
Copy matrices $A$, $W$ and $H$ from the host to the GPU\;
\Begin{
	\Initialize{$H \leftarrow \text{(small random numbers)}$}\;
	\For{$iter\leftarrow 1$ \KwTo $T$ $Step=1$}{
		Update $W$ using $\omega_i^*= (H_{\Omega _i}^T H_{\Omega _i} + \lambda I )^{-1} H^T a_i$;
		\tcp*[h]{Host Sync}\;
		Update $H$ using $h_j^*= (W_{\Omega _j}^T W_{\Omega _j} + \lambda I )^{-1} W^T a_j$;
		\tcp*[h]{Host Sync}\;
	}
	Copy matrices $W$ and $H$ from GPU to host\;	
}
\caption{GPU-ALS}\label{alg:ALSP_CUDA_ALG}
\end{algorithm}%\DecMargin{1em}
} % end small

\section{Parallel CCD++}

In the CCD++ algorithm, each solution is obtained by
alternately updating $W$ and $H$. When $v$ is constant, each variable $u_i$ is updated
independently (eq. \ref{eq:minUi_CCDFeauture}). Therefore, the update of $u$ can be made by
several processing cores. 

Given a computer with $p$ cores, we define the partition of the
row indexes of $W,\{1,\ldots,m\}$ as $S=S_1,\ldots,S_p$. Vector $u$ is
decomposed in $p$ vectors $u^1,u^2,\ldots,u^p$, where $u^r$ is the
sub-vector of $u$ corresponding to $S_r$. When the matrix $W$ is
uniformly split in parts $\lvert S_1 \rvert=\lvert S_2 \rvert = \ldots
= \lvert S_p \rvert = \frac{m}{p}$, there is a load balancing problem
due to the variation of the size of the row vectors contained in $W$.
In this case, the exact amount of work for each $r$ core to
update $u^r$ is given by $\sum_{i \in S_r} 4 \lvert \Omega_i
\rvert$~\cite{kais-pmf}. Therefore, different cores have different
workloads.  This is one of the limitations of this algorithm. It can
be overcome using dynamic scheduling, which is offered by most
parallel processing libraries (e.g.
OpenMP~\cite{chandra2001parallel}).

For each subproblem, each core $r$ builds $\hat{R}$ with,
{\small
	\begin{equation} \label{eq:computeCCD_feautureR_p}
		\hat{R}_{ij} \leftarrow R_{ij}+\bar{\omega} _{ti}\bar{h} _{tj}, \forall (i,j) \in \Omega_{S_r} ,
	\end{equation}
}
\noindent
where $\Omega_{S_r} = \cup _{i \in S_r} \{ (i,j) : j \in \Omega_i$. Then, for each core $r$ we have,
{\small
	\begin{equation} \label{eq:minUi_CCDFeauture_p}
		u_i \leftarrow \frac{\sum\limits_{j \in \Omega_i} \hat{R}_{ij} v_j}{\lambda + \sum\limits_{j \in \Omega_i} v^2_j}, \forall i \in S_r.
	\end{equation}
}
The update of $H$ is analogous to the one of $W$ in
(\ref{eq:minUi_CCDFeauture_p}). For $p$ cores the row indexes
of $H,\{1,\ldots,n\}$ are partitioned into $G=G_1,\ldots,G_p$. So, for
each core $r$ we have,
{\small
	\begin{equation} \label{eq:minVj_CCDFeauture_p}
		v_j \leftarrow \frac{\sum\limits_{i \in \bar{\Omega}_j} \hat{R}_{ij}u_j}{\lambda + \sum\limits_{i \in \bar{\Omega}_j} u^2_i}, \forall j \in G_r.
	\end{equation}
}
Since all cores share the same memory, no communication is
needed to access $u$ and $v$. After obtaining $(u^*,v^*)$, the update
of $R$ and $(\bar{\omega}_t^r,\bar{h}_t^r)$ is also implemented in
parallel by the $r$ cores as follows.
{\small
	\begin{equation} \label{eq:updateOmegaH_CCDPP_p}
		(\bar{\omega}_t^r,\bar{h}_t^r) \leftarrow (u^r,v^r),
	\end{equation}	

	\begin{equation} \label{eq:updateR_CCDPP_p}
		R_{ij} \leftarrow \hat{R}_{ij} - \bar{\omega}_{ti} \bar{h}_{tj}, \forall (i,j) \in \Omega_{S_r}.
	\end{equation}	
}
Algorithm~\ref{alg:ccdppCCD_Parallel} summarizes the parallel CCD operations.
{\small
%\IncMargin{1em}
\SetAlFnt{\small}
\begin{algorithm}[htb]
\SetKwData{Left}{left}\SetKwData{This}{this}\SetKwData{Up}{up}
\SetKwFunction{Union}{Union}\SetKwFunction{FindCompress}{FindCompress}
\SetKwInOut{Input}{input}\SetKwInOut{Output}{output}
\SetKwFunction{Initialize}{initialize}
\Input{$A, W, H, \lambda, k, T$}
	\Initialize{$W\leftarrow 0, R\leftarrow A$}\;	
	\For{$iter\leftarrow 1$ $\ldots$ $Step=1$}{		
		\For{$t\leftarrow 1$ \KwTo $k$ $Step=1$}{
			in parallel, build $\hat{R}$ split by $r$ cores using (\ref{eq:computeCCD_feautureR_p})\;
			\For{$inneriter\leftarrow 1$ \KwTo $T$ $Step=1$}{				
				in parallel, update $u$ with $r$ cores using (\ref{eq:minUi_CCDFeauture_p})\;
				in parallel, update $v$ with $r$ cores using (\ref{eq:minVj_CCDFeauture_p})\;
			}	
			in parallel, update  $(\bar{\omega}_t^r,\bar{h}_t^r)$ using (\ref{eq:updateR_CCDPP_p})\;	
			in parallel, update  $R$ using (\ref{eq:updateR_CCDPP_p})\;
		}	
	}

\caption{Multi-core version of CCD++~\cite{kais-pmf}}\label{alg:ccdppCCD_Parallel}
\end{algorithm}%\DecMargin{1em}
} %end small
%\newpage
\section{CCD++ in CUDA}
Our CUDA implementation of the CCD++ algorithm uses
explicit memory management. It is inspired by the parallel version of
CCD++ found in LIBPMF (Library for Large-scale Parallel
  Matrix Factorization). This is an open source library for
Linux~\cite{kais-pmf}.  LIBPMF is implemented in
C++ for multi-core environments with shared memory. The
parallel version uses the OpenMP
library~\cite{chandra2001parallel}. It employs double precision
values. Our version uses floats because GPUs are faster when floats
are used.
	%http://www.cs.utah.edu/~vishayv/takagi_cuda_paper.pdf
{\small	
\SetAlFnt{\small}
\begin{algorithm}[htb]
\SetKwData{Left}{left}\SetKwData{This}{this}\SetKwData{Up}{up}
\SetKwFunction{Union}{Union}\SetKwFunction{FindCompress}{FindCompress}
\SetKwInOut{Input}{input}\SetKwInOut{Output}{output}
\SetKwFunction{Initialize}{initialize}
\Input{$A, W, H, \lambda, k, T$}
	\Initialize{$W\leftarrow 0, R\leftarrow A$}\;
	Allocate memory on GPU for matrices $A$ and $R$ and for vectors $u$ and $v$\;
	Copy matrices $A$ and $R$ from host to GPU\;
	\For{$iter\leftarrow 1$ \KwTo $T$ $Step=1$}{		
		\For{$t\leftarrow 1$ \KwTo $k$ $Step=1$}{
			$u \leftarrow \bar{\omega}_t$ and $v \leftarrow \bar{h}_t$\;
			Copy vectors $u$ and $v$ from host to GPU\;
			call {\em kernel} to update $\hat{R}$ on GPU using (\ref{eq:computeCCD_feautureR_p})\;
			\For{$inneriter\leftarrow 1$ \KwTo $T$ $Step=1$}{				
				update $u$ and $v$ on GPU using (\ref{eq:minUi_CCDFeauture_p}) and (\ref{eq:minVj_CCDFeauture_p})\;
			}	
			Copy vectors $u$ and $v$ from GPU to host\;
			$\bar{\omega}_t \leftarrow u$ and $\bar{h}_t \leftarrow v$\;
			update $\hat{R}$ on GPU using (\ref{eq:updateR_CCDPP_p})\;
		}	
	}
\caption{CCD++ GPU Implementation}\label{alg:ccdppALG_CUDA}
\end{algorithm}%\DecMargin{1em}	
} %end small

Algorithm~\ref{alg:ccdppALG_CUDA} shows our implementation of the
CCD++ for the GPUs.

We use the same stream in all copies from host to
device, device to host and for
{\em kernels}. Therefore, each of the operations is always blocking
with respect to the main thread in the host.

%\newpage

\section{Materials and Methods}

\noindent
We performed our experiments using two operating systems: Windows 8.1 pro x64
and Linux fedora 20. The CUDA versions for these two systems can vary
greatly in performance. The hardware used is described as follows:
	\textbf{GPU:} Gainward GeForce GTX 580 Phantom, $\approx$ $\$600$,
			with total dedicated memory 3GB GDDR5 and
			512 CUDA Cores; \textbf{Processors:} 2 $\times$
          Intel\textsuperscript{\textregistered}
          Xeon\textsuperscript{\textregistered} X5550, $2\times \$999
          \approx \$1998 $, with 24GB of RAM (6 $\times$ 4GB HYNIX HMT151R7BFR4C-H9);
		\textbf{Motherboard:} Tyan S7020WAGM2NR.
%		\item \textbf{hard disk for Windows 8.1 pro x64:} Samsung SSD 840 PRO 256GB
%		\item \textbf{hard disk for Linux:} Samsung SSD 840 PRO 128GB				
%	\end{itemize}	

All experiments use the Netflix
  dataset (100,480,507 ratings that 480,189 users gave to 17,770
movies). Our qualitative evaluation metric is the root mean squared error
(RMSE) produced on the probe
  data generated by the model. Our quantitative measure is the
speedup (how fast it is the parallel implementation related to
the sequential, calculated as the sequential execution time divided by the
parallel execution time). 

Ideally, we needed a secondary GPU with dedicated memory, but this was
not possible. In our GPU, the memory is shared with the display
memory. We used 16 blocks of 512 threads in our experiments.

The parameters used by both CCD++ and ALS are $k = 5$,
$\lambda = 0.1 $ and $T = 15$. These were selected according to an
empirical selection. Lower values of $k$ give better speedups for the
GPU implementation, while a variation of the $k$ values does not
impact the multi-core implementation. Higher values of $k$ also implies
that more data will be copied to the GPU memory, which is not advisable. 

We performed our experiments with two versions of the CCD++, one using
float (single decimal precision) and the other using doubles (double
decimal precision), in order to evaluate how the GPU would behave with
both kinds of numeric types.

All experiments for CCD++ resulted on RMSE
equals to $0.94$ and for ALS resulted in RMSE equals to $0.97$.

\section{Results and Discussion}

%\subsection{CCD++}
Table~\ref{table:OMP_libpmf_speedup_netflix} shows the performance of
the original CCD++ (using the library libpmf) on the multi-core machine
with Linux and Windows, running the Netflix benchmark, using the
original double decimal precision (C double). The speedups achieved in
Windows are higher than in Linux, but this was expected, since the
base execution of Windows (717.3 s for 1 thread) is higher than the
Linux (521.5 s). The maximum speedup achieved is 4.4 with 32
threads. %These results are similar to the ones reported by Yu {\it et
 % al.}~\cite{kais-pmf}. %ICD: André, poderia
                                %confirmar se é isso mesmo para este
                                %benchmark?
\begin{table}[!htb]
\caption{LIBPMF with double in OMP.}
\centering
{\small
\begin{tabular}{ | p{2cm} | l | l |}		
	\hline
	\multicolumn{3}{|c|}{OS: Linux} \\
	\hline
	Test   & Execution time & Speedup \\ \hline
	1 thread & $\pm 521.512s$ &    \\ \hline
	2 threads & $\pm 316.701s$ & $1.6$   \\ \hline
	8 threads & $\pm 136.2s$ & $3.8$ \\ \hline
	16 threads& $\pm 126.81s$ & $4.1$ \\ \hline
	32 threads& $\pm 136.023s$ & $3.8$ \\ \hline 
	\hline
	\multicolumn{3}{|c|}{OS: Windows 8.1 pro x64} \\
	\hline	
	Test    & Execution time & Speedup \\ \hline
	1 thread & $\pm 717.307s$ &    \\ \hline
	2 threads & $\pm 407.873s$ & $1.8$   \\ \hline
	8 threads & $\pm 179.499s$ & $4.0$ \\ \hline
	16 threads& $\pm 166.746s$ & $4.3$ \\ \hline
	32 threads& $\pm 161.48s$ & $4.4$ \\ \hline 					
\end{tabular}	
} % end small
\label{table:OMP_libpmf_speedup_netflix}		
\end{table}

Table~\ref{table:OMP_cuda_libpmf_speedup_netflix} shows the same
experiments, but now with our version of CCD++, that uses a single
decimal precision. The results are exactly the same in terms of RMSE,
but the performance is highly benefited by the numeric data type in
this case. By using floats, instead of doubles, we reach speedups of
9.5 (at 32 threads), which is more than twice the speedup achieved
with the version that used a double numeric representation. Note that
the original libpmf uses doubles instead of floats. We could achieve
even better speedups than they reported, by just using single
precision data. The use of float or double did not affect much the
Linux implementations, but it considerably affected the Windows
implementations. % André, alguma razão
                                % para esta difeença entre floats and
                                % doubles em Windows? A libpmf usa
                                % floats ou doubles originalmente?

% André, por que a libpmf é implementada
% com doubles? Alguma razão especial?

Again, with this version, the Windows implementation achieves higher
speedups than Linux. This was expected, since the base execution
time for 1 thread is much higher for Windows. 

\begin{table}[!htb]
\caption{CCD++ with float in OMP and CUDA.}
\centering
{\small
\begin{tabular}{ | p{2cm} | l | l |}		
	\hline
	\multicolumn{3}{|c|}{OS: Linux} \\
	\hline		
	Test   & Execution time & Speedup \\ \hline
	1 thread & $\pm 528.538s$ &    \\ \hline
	2 threads & $\pm 309.707s$ & $1.7$ \\ \hline
	8 threads& $\pm 111.968s$ & $4.7$ \\ \hline
	16 threads& $\pm 98.1266s$ & $5.3$ \\ \hline
	32 threads& $\pm 99.8027s$ & $5.2$ \\ \hline
	CUDA & $\pm\textbf{168.109s}$ & $\textbf{3.1}$ \\ \hline
	\hline
	\multicolumn{3}{|c|}{OS: Windows 8.1 pro x64} \\
	\hline		
	Test & Execution time & Speedup \\ \hline
	1 thread & $\pm 1252.35s$ &    \\ \hline
	2 threads & $\pm 540.973s$ & $2.3$ \\ \hline
	8 threads& $\pm 181.501s$ & $6.9$ \\ \hline
	16 threads& $\pm 131.881s$ & $9.5$ \\ \hline
	32 threads& $\pm 131.661s$ & $9.5$ \\ \hline
	CUDA & $\pm\textbf{84.7718s}$ & $\textbf{14.8}$ \\ \hline				
	\end{tabular}
} % end small
\label{table:OMP_cuda_libpmf_speedup_netflix}			
\end{table}	

But our best results are for the CUDA implementation. We obtained
a speedup of 14.8 just using the GPU running our implementation of the
CCD++ in Windows. We managed to surpass the performance of a machine
that costs more than twice as much as a GPU card, showing that these
architectures have a great potential for the implementation of recommender
systems based on matrix factorization.

%\newpage

\subsection{ALS}

	\begin{table}[!htb]
	\caption{ALS with float in OMP
          and  CUDA.}
	\centering
{\small	
	\begin{tabular}{ | p{2cm} | l | l |}		
		\hline
		\multicolumn{3}{|c|}{OS: Linux} \\
		\hline		
		Test   & Execution time & Speedup \\ \hline
		1 thread & $\pm 429.539s$ &    \\ \hline
		2 threads & $\pm 224.99s$ & $1.9$ \\ \hline
		8 threads& $\pm 93.8716s$ & $4.6$ \\ \hline
		16 threads& $\pm 98.3057s$ & $4.3$ \\ \hline
		32 threads& $\pm 95.8294s$ & $4.5$ \\ \hline
		CUDA & $\pm\textbf{98.71s}$ & $\textbf{4.4}$ \\ \hline
		\hline
		\multicolumn{3}{|c|}{OS: Windows 8.1 pro x64} \\
		\hline		
		Test    & Execution time & Speedup \\ \hline
		1 thread & $\pm 665.74s$ &    \\ \hline
		2 threads & $\pm 355.144s$ & $1.9$ \\ \hline
		8 threads& $\pm 158.912s$ & $4.2$ \\ \hline
		16 threads& $\pm 121.667s$ & $5.5$ \\ \hline
		32 threads& $\pm 122.121s$ & $5.5$ \\ \hline
		CUDA & $\pm\textbf{107.214s}$ & $\textbf{6.2}$ \\ \hline	
		\end{tabular}
 } % end small
	\label{table:OMP_cuda_ccdpp_speedup_netflix}			
	\end{table}		

We also implemented the ALS algorithm in CUDA and results are
presented in Table~\ref{table:OMP_cuda_ccdpp_speedup_netflix} for
comparison. In this table we show execution times and speedups for the
multi-core version and for the GPU. Once more the multi-core version presents
better speedups with a higher number of threads, for the Windows
environment. % André, como é que estes resultados multicore comparam
             % com a versão CUDA ALS na literatura?

%\balance	
\section{Conclusions}
We showed the advantage of using GPUs to implement recommender systems
based on matrix factorization algorithms. Using a benchmark popular in
the literature, Netflix, we obtained maximum speedup of 14.8, better
than the best speedup reported in the literature. % André, precisa ver
                                % qual é o melhor speedup conseguido
                                % em GPUs para o ALS

The advantages of using a CUDA implementation over a multi-core server
are: lower energy consumption, lower price and the ability of leaving
the main host or other cores to be used by other
tasks. Currently, almost every computer comes with PCI slots that can
be used to install a GPU or various GPUs. Thus, it is relatively
simple to expand the computational capacity of an existing hardware. 

We plan to perform more tests with our algorithms on more recent GPUs and on larger datasets. One potential problem of GPUs is their memory limitation. Therefore, one path
to follow is to implement efficient memory management mechanisms
capable of dealing with bigger data. Another track we would like to
follow is to implement a load balancing mechanism to these algorithms.

%% Neste momento a NVIDIA det\'em duas gamas distintas de placas
%% adequadas para processamento CUDA. A gama mais econ\'omica \'e a GTX,
%% uma GTX topo de gama consegue um processamento equivalente \`as
%% melhores da gama Tesla que \'e a gama mais cara de todas. A diferen\c
%% ca da gama Tesla est\'a em ser dedicada para servidores e apenas para
%% processamento CUDA, mas o que a torna especial \'e a garantia de ser
%% 100\% \`a prova de erros por deter mem\'oria ECC protected. Mesmo
%% assim quando os erros pequenos n\~ao afetam os resultados, tal como no
%% caso dos sistemas de recomenda\c c\~ao, uma placa da gama GTX \'e
%% suficiente, note-se que apenas n\~ao \'e dada garantia de ser 100\%
%% \`a prova de erro e o mesmo acontece em multicore nos computadores
%% pessoais sem mem\'oria ECC. Existe ainda a gama Quadro, mas essa \'e
%% especificamente dedicada a processamento e tratamento de v\'ideo. Por
%% isso dependendo das caracter\'isticas do algoritmo, gastando o mesmo
%% dinheiro num ou dois processadores, \'e poss\'ivel comprar uma placa
%% que consiga uma performance superior. Isto acontece nos casos em que o
%% algoritmo aproveita ao m\'aximo as capacidades da tecnologia, no
%% entanto isso nem sempre \'e poss\'ivel.

%% Neste artigo demonstra-se que \'e poss\'ivel obter speedup utilizando CUDA em sistemas de recomenda\c c\~ao baseados em factoriza\c c\~ao de matrizes.

%ACKNOWLEDGMENTS are optional
\section{Acknowledgments}
National Funds through the FCT - Funda\c{c}\~ao para a Ci\^encia e a Tecnologia (proj. FCOMP-01-0124-FEDER-037281).
%%This work is supported by grant (PTDC/ EEI-SII/ 2094/
%%2012) and ... NEED TO BE COMPLETED HERE!

\bibliographystyle{abbrv}
{
\small

\vphantom{1cm}

\bibliography{sigproc}  % sigproc.bib is the name of the Bibliography in this case
} % end small
\end{document}